\documentclass[12pt]{article} 
\usepackage{epsfig} \usepackage{amsfonts,amssymb,amsmath}
\usepackage{subfigure}

\newcommand{\ga}{\ensuremath{\gamma}}

\renewcommand{\t}{\ensuremath{\tau}}

\newcommand{\del}{\ensuremath{\partial}}

 \newcommand{\be}{\begin{equation}}
\newcommand{\ee}{\end{equation}} 
\newcommand{\ba}{\begin{eqnarray}} \newcommand{\ea}{\end{eqnarray}}

\newcommand{\lab}[1]{\label{#1}}  
\newcommand{\bib}[1]{\bibitem{#1}}

 \newcommand{\nc}{\newcommand}

\def\ie{\textit{i.e. }} \def\mn{{\mu\nu}} \def\tt{\textrm}

\nc{\aaa}[3]{{\ Astron.\ Astroph.\ }{{\bf #1},({#2}){#3}}}
\nc{\advp}[3]{{Adv.\ in\ Phys.\ }{{\bf #1}({#2}){#3}}}
\nc{\apl}[3]{{Appl. Phys. Lett. }{{\bf #1}{(#2)}{#3}}}
\nc{\apj}[3]{{Astrophys.\ J.\ }{{\bf #1} {(#2)} {#3}}}
\nc{\apjl}[3]{{Astrophys.\ J.\ Lett.\ }{{\bf #1} {(#2)} {#3}}}
\nc{\app}[3]{{\ Astrop.\ Phys..\ }{{\bf #1}, {(#2)} {#3}}}
\nc{\asp}[3]{{\ Astropart.\ Phys.\ }{{\bf #1} {(#2)} {#3}}}

\nc{\cmp}[3]{{  Comm.\ Math.\ Phys.\ }{{ \bf #1} {(#2)} {#3}}}
\nc{\cqg}[3]{{  Class.\ Quant.\ Grav.\ }{{\bf #1} {(#2)} {#3}}}
\nc{\epl}[3]{{  Europhys.\ Lett.\ }{{\bf #1} {(#2)} {#3}}}
\nc{\ijmp}[3]{{ Int.\ J.\ Mod.\ Phys.\ }{{\bf #1} {(#2)} {#3}}}
\nc{\ijtp}[3]{{ Int.\ J.\ Theor.\ Phys.\ }{{\bf #1} {(#2)} {#3}}}
\nc{\jhep}[3]{{ JHEP\ }{{\bf #1} {(#2)} {#3}}} \nc{\jmp}[3]{{  J.\
Math.\ Phys.\ }{{ \bf #1} {(#2)} {#3}}} \nc{\jpa}[3]{{  J.\ Phys.\ A\
}{{\bf #1} {(#2)} {#3}}} \nc{\jpc}[3]{{  J.\ Phys.\ C\ }{{\bf #1}
{(#2)} {#3}}} \nc{\jpg}[3]{{ J.~Phys.~G:~Nucl.~Part.~Phys.~}{{\bf #1}
{(#2)} {#3}}} \nc{\jap}[3]{{ J.\ Appl.\ Phys.\ }{{\bf #1} {(#2)}
{#3}}} \nc{\jpsj}[3]{{ J.\ Phys.\ Soc.\ Japan\ }{{\bf #1} {(#2)}
{#3}}} \nc{\lmp}[3]{{ Lett.\ Math.\ Phys.\ }{{\bf #1} {(#2)} {#3}}}
\nc{\lncim}[3]{{ Lett.\ Nuov.\ Cim.\ }{{\bf #1} {(#2)} {#3}}}
\nc{\mpl}[3]{{ Mod.\ Phys.\ Lett.\ }{{\bf #1} {(#2)} {#3}}}
\nc{\nat}[3]{{  Nature \ }{{\bf #1} {(#2)} {#3}}} \nc{\ncim}[3]{{
Nuov.\ Cim.\ }{{\bf #1} {(#2)} {#3}}} \nc{\npb}[3]{{ Nucl.\ Phys.\
}{{\bf B#1} {(#2)} {#3}}} \nc{\pr}[3]{{ Phys.\ Rev.\ }{{\bf #1} {(#2)}
{#3}}} \nc{\pra}[3]{{  Phys.\ Rev.\ }{{\bf A#1} {(#2)} {#3}}}
\nc{\prb}[3]{{  Phys.\ Rev.\ }{{\bf B#1} {(#2)} {#3}}} \nc{\prc}[3]{{
Phys.\ Rev.\ }{{\bf C#1} {(#2)} {#3}}} \nc{\prd}[3]{{  Phys.\ Rev.\
}{{\bf D#1} {(#2)} {#3}}} \nc{\prl}[3]{{ Phys.\ Rev.\ Lett.\ }{{\bf
#1} {(#2)} {#3}}} \nc{\plb}[3]{{  Phys.\ Lett.\ }{{\bf B#1} {(#2)}
{#3}}} \nc{\prep}[3]{{ Phys.\ Rep.\ }{{\bf #1} {(#2)} {#3}}}
\nc{\prsl}[3]{{ Proc.\ R.\ Soc.\ London\ }{{\bf #1} {(#2)} {#3}}}
\nc{\ptp}[3]{{  Prog.\ Theor.\ Phys.\ }{{\bf #1} {(#2)} {#3}}}
\nc{\ptps}[3]{{ Prog\ Theor.\ Phys.\ suppl.\ }{{\bf #1} {(#2)} {#3}}}
\nc{\physa}[3]{{ Physica\ A\ }{{\bf #1} {(#2)} {#3}}} \nc{\physb}[3]{{
Physica\ B\ }{{\bf #1} {(#2)} {#3}}} \nc{\phys}[3]{{ Physica\ }{{\bf
#1} {(#2)} {#3}}} \nc{\rmp}[3]{{ Rev.\ Mod.\ Phys.\ }{{\bf #1} {(#2)}
{#3}}} \nc{\rpp}[3]{{ Rep.\ Prog.\ Phys.\ }{{\bf #1} {(#2)} {#3}}}
\nc{\sjnp}[3]{{ Sov.\ J.\ Nucl.\ Phys.\ }{{\bf #1} {(#2)} {#3}}}
\nc{\jetp}[3]{{ JETP\ }{{\bf #1} {(#2)} {#3}}} \nc{\yf}[3]{{ Yad.\
Fiz.\ }{{\bf #1} {(#2)} {#3}}} \nc{\zetp}[3]{{ Zh.\ Eksp.\ Teor.\
Fiz.\ }{{\bf #1} {(#2)} {#3}}} \nc{\zp}[3]{{ Z.\ Phys.\ }{{\bf #1}
{(#2)} {#3}}} \nc{\zpc}[3]{{ Z.\ Phys.\ C\ }{{\bf #1} {(#2)} {#3}}}
\nc{\ibid}[3]{{\sl ibid.\ }{{\bf #1} {#2} {#3}}}

\begin{document}


\begin{center}
{\Large \bf Braneworld Isotropization and Magnetic Fields}
\end{center}
\vskip 1cm \renewcommand{\thefootnote}{\fnsymbol{footnote}}

\centerline{\bf
Gustavo Niz$^{a}$\footnote{gustavo.niz@nottingham.ac.uk}, Antonio
Padilla$^{a}$\footnote{antonio.padilla@nottingham.ac.uk}, Hari. K. Kunduri$^{a,b}$\footnote{h.k.kunduri@damtp.cam.ac.uk},}
\vskip .5cm

\centerline{$^a$ \it School of Physics and Astronomy} \centerline{\it
University of Nottingham, NG7 2RD, UK} \vskip .5cm
\centerline{$^b$ \it DAMTP, Centre for Mathematical Sciences} \centerline{\it
University of Cambridge, CB3 0WA, UK}

\setcounter{footnote}{0} \renewcommand{\thefootnote}{\arabic{footnote}}


\begin{abstract}
We consider a magnetic Bianchi I braneworld, embedded in between two Schwarzschild-AdS spacetimes, boosted equal amounts in opposite directions and compare them to the analagous solution in four-dimensional General Relativity. The efficient dissipation of anisotropy on the brane is explicitly demonstrated, a process we dub braneworld isotropization. From the bulk point of view, we attribute this to anisotropic energy being carried into the bulk by hot gravitons leaving the brane. From the brane point of view this can be interpreted in terms of the production of particles in the dual CFT. We explain how this result  enables us to gain a better understanding  of the behaviour of anisotropic branes already studied in the literature.  We also show how there is evidence of particles being over-produced, and comment on how this may ultimately provide a possible observational signature of braneworlds. 
\end{abstract}

\newpage

\section{Introduction}
The most recent data from WMAP suggests that the Cosmic Microwave
Background (CMB) is isotropic to within one part in
$10^5$~\cite{wmap}. Why is our universe {\it so} isotropic? This has
been an important question in cosmology ever since the CMB's discovery
back in the mid 1960s~\cite{cmb}. The obvious way to approach this is
to assume that the universe began in a highly {\it an}isotropic state,
and ask what dynamical mechanism caused the universe to shed nearly
all its anisotropy, leaving behind the highly symmetric state we
observe today.

The most popular mechanism is, of course,
inflation~\cite{inflation}. In inflation, almost all anisotropy was
diluted away during a period of accelerated expansion in the early
universe. The accelerated expansion was driven by a scalar field
rolling slowly down its potential,  closely resembling a positive
cosmological constant. Indeed, it can be shown that in the presence of
a positive cosmological constant, all but one of the Bianchi
models\footnote{The exception is  Bianchi IX.}, describing a
homogeneous but anisotropic cosmology, evolve asymptotically to an
isotropic de Sitter solution~\cite{wald}. However, there is still an
ongoing discussion on how fine-tunned the initial conditions should be
to get a sufficiently flat, homogeneous and isotropic universe after
inflation (see \cite{carrol} and references therein).

It is often easy to forget that before inflation, plenty of other
``isotropization''mechanisms were put forward, with varying degrees of
success. The earliest was probably the ``phoenix'', or
``oscillatory'',  universe~\cite{phoenix}, which expands for a while,
recollapses towards a bounce, and then starts to expand again. As
entropy builds up after each oscillation, it was argued that the
period of oscillation increased, and the  universe approached an
isotropic, homogeneous state of thermal equilibrium. Although these
ideas ran into problems with  inevitable singularities~\cite{sings},
they have been revived recently with the advent of the cyclic
universe~\cite{cyclic} and a possible resolution to its
singularity~\cite{niz}.

Another means of damping anisotropy is particle production due to
quantum effects at very early times~\cite{pp}. Energetically, this can
be understood as anisotropy being converted into thermal
radiation. If, for example, the universe were contracting along one
direction, but expanding along the others, the momenta of virtual
particles would be blueshifted along the contracting direction. This
would make the virtual particles more likely to materialize as real
particles, drawing energy from the contraction, and lowering the
amount of anisotropy.  However, if we assume an arbitrarily large
amount of initial anisotropy, a huge amount of thermal radiation would
need to be produced to account for the levels of isotropy seen in the
CMB, and as a result, the photon-baryon ratio would massively exceed
its observed value~\cite{bm}. Particle production alone cannot,
therefore, account for enough dissipation, and we typically assume it
to be one of many contributing factors, including neutrino
viscosity~\cite{viscosity}, and, inevitably, inflation.

In this paper, we point out another  mechanism for dissipating
anisotropy: {\it braneworld isotropization}. Consider a
Randall-Sundrum type braneworld~\cite{rs}, and assume it is highly
anisotropic. As the brane evolves, the anisotropic energy on the brane
leaks into the bulk, and the brane becomes rapidly more isotropic,
whilst the bulk becomes { less} homogeneous and/or { less}
isotropic. To get a feel for how this works consider the projection of
the Einstein equation onto a Randall-Sundrum braneworld~\cite{sms},
schematically given by 
\be 
G_\mn(\gamma)=8 \pi G_4
T_\mn+G_5^2(T_\mn)^2-E_\mn, 
\ee 
where $G_4$ and $G_5$ are the brane and
bulk Newton's constants respectively, and  $T_\mn$ is the
energy-momentum tensor for matter  on the brane.  Note that we have a
non-local piece, $E_\mn$, which is the electric part of the  bulk Weyl
tensor, projected onto the brane.  The ``anisotropic energy density''
on the brane, in the sense described in~\cite{viscosity},   is  stored
in the local geometrical piece, $G_\mn$, and as the brane evolves,
this energy is transferred into the bulk through the bulk Weyl term,
$E_\mn$.

The contribution of the bulk Weyl term is often best understood
holographically (see, for example~\cite{bwholo3}) using the braneworld
version of the AdS/CFT correspondence~\cite{adscft, bwholo1}. This
states that {\it  Randall-Sundrum braneworld gravity is dual to a CFT,
with a UV cut-off, coupled to gravity in $3+1$ dimensions}. It follows
that a {\it classical} source on the brane behaves like a {\it
quantum} source in $3+1$ dimensons, in the sense that it has been
dressed with quantum corrections from the CFT~\cite{bwholo2}.  The
anisotropy dissipation process we have just described can now be
interpreted as particle production in the CFT. Energy is drawn from
the anisotropy to fuel the production of CFT particles, leading to
isotropization on the brane.

There is already evidence in the literature for braneworld
isotropization. Consider, for example, the quest to find anisotropic
geometries supporting a perfect fluid. This is easy enough in
four-dimensional General Relativity, the Bianchi or Kantowski-Sachs
metrics being good examples~\cite{kantowski, exacsols}. In brane
cosmology, full solutions have not been so easy to 
find~\cite{maar1, orsaylot, frolov}, and only a rather contrived set
of kasner-like solutions are known exactly~\cite{orsaylot,
frolov}. Indeed, one can prove that if the bulk is static, an
anisotropic brane cannot support a perfect fluid~\cite{orsaylot}. This
result can now be  easily understood: particle production in the CFT
leads to  dissipation of anisotropy over time. From the $5D$ point of
view, the leaking of anisotropy into the bulk prevents it from being
static. Another interesting example of braneworld isotropization lies
in the study of anisotropy dissipation in braneworld
inflation~\cite{maar2}. There it was noticed that inflation begins
sooner than it would in $4D$ General Relativity. We attribute this to
particle production in the CFT drawing its energy from the anisotopy,
aswell as the kinetic energy of the inflaton. 

In this paper, we demonstrate braneworld isotropization explicitly by
means of a concrete example. We generate the initial anisotropy on a Bianchi I braneworld by means of a constant magnetic field. Unlike much of the existing literature on
anisotropic braneworlds  the full bulk solution is known,
corresponding to the planar limit of two boosted Schwarzschild-AdS black hole
spacetimes, cut and pasted together to form the brane in the usual way. Because the brane embedding is highly non-trivial when viewed in global coordinates, it is  convenient to work in a boosted coordinate system on either side of the brane, so that the bulk metric resembles the planar limit of two particular Myers-Perry-AdS spacetimes~\cite{MP,HHT}. The boosts are equal and opposite on either side of the brane, so one may think of the corresponding Myers-Perry solutions as having equal and opposite angular momentum. We then track the
brane's evolution, paying particular attention to the anisotropic
shear, and compare it to the corresponding scenario with the same
source in four-dimensional General Relativity. As expected, the shear
anisotropy dissipates far quicker in the Randall-Sundrum braneworld
scenario, than in $4D$ GR.

A study of magnetic fields on a braneworld is also important in its own right. One of
the great puzzles in  astrophysics concerns the origin of large
magnetic fields in galaxies and galaxy clusters (for an excellent
review of magnetic fields in the early universe,
see~\cite{magrev}). It has been argued that these fields were
generated by a galactic dynamo process~\cite{dynamo}, sourced by a
smaller pre-galactic ``seed'' field. However, one still has to account
for the origin of the seed field, and even then the efficiency of this
process has been brought into question~\cite{dynamobad}. We are
naturally led to consider the possibility that galatic fields result
from large scale primordial fields left over from the big bang.  In
the standard cosmology, one can place bounds on the size of a large
scale primordial magnetic field from nucleosynthesis ($B \lesssim
10^{-7}$ G)~\cite{BBNbound}, and from CMB temperature fluctuations ($B
\lesssim 10^{-9}$ G)~\cite{CMBbound}. Given that the CMB bound is the
stronger, we might speculate that this bound is
weakened in a braneworld context, due to isotropization. 

Magnetic fields on branes have been considered in the
past~\cite{magbrane, ecc}, although a full knowledge of the bulk has
been absent. Whilst it is true that one can certainly learn a lot {\it
without} full knowledge of the bulk, we believe it is a dangerous game
to play. It is not at all obvious that a bulk solution that is regular
near the brane evolves into something regular far from the brane. A
good example of this is the black string solution~\cite{bwbh} which is
regular near the brane, but singular on the AdS horizon. 

The rest of this paper is organised as follows. In
section~\ref{sec:4dgr}, we review the Bianchi I solution for a
constant magnetic field in four-dimensional General Relativity. For
completeness, and in keeping with supernovae
observations~\cite{supernova}, we will allow for  a positive
cosmological constant that can be set to zero if necessary. In
section~\ref{sec:bw}, we introduce a certain Myers-Perry-AdS black hole in
five dimensions, and take the planar limit, so that we end up with planar Schwarzschild-AdS in boosted coordinates. We then cut and paste two equal and
oppositely boosted black holes onto one another, in the usual way, in order to
form a Bianchi I braneworld, supported by brane tension, and a
constant magnetic field.  In section~\ref{sec:compare} we compare the
two scenarios, with special emphasis on the  dissipation of
anisotropy, and the effect of particle production in the CFT. Finally,
in section~\ref{sec:discuss}, we summarize our results.

Our conventions follow the Landau-Lifshitz notation for the curvature
tensors, and we use a $(-,+,...,+)$ signature for the metric.
\section{A homogeneous magnetic field in $4D$ GR} \lab{sec:4dgr}
Let us begin with a review of homogeneous magnetic fields in
four-dimensional General Relativity. As we have already mentioned, a
large scale magnetic field of this type could be left over from the
big bang, and  may provide the seed for fields seen in galaxies and
galaxy clusters.  In particular, we will study four-dimensional
Einstein-Maxwell theory in the presence of a cosmological constant,
$\Lambda_4$. The cosmological constant has been included for both
completeness and phenomenological reasons~\cite{supernova}. Since we
will ultimately be comparing rates of isotropization in $4D$ GR and on
the brane, we might be concerned that the cosmological constant will
play the dominant role in each case and make comparisons difficult. As
Wald proved~\cite{wald}, this is indeed the case, but we can account
for it, and in any case, we can always set the cosmological constant
to zero if we so desire.

The action for the system is \be\lab{action0} S=\int
d^{4}x\sqrt{-\gamma}\left[\frac{1}{16\pi
G_4}\left(R-2\Lambda_4\right)- \frac{1}{4}F^{\mu\nu}F_{\mu\nu}\right],
\ee where $G_4$ is Newton's constant in four dimensions and
$\gamma_{\mu\nu}$ is the metric of the spacetime spanned by the
coordinates $(t,x,y,z)$. We will assume that the cosmic plasma is a
perfect conductor, and so any primordial electric field is screened by
the plasma in the early universe. We are therefore only left with a
primordial magnetic field, which we take to be homogeneneous, and
directed along (say) the $x$ direction. The field is assumed to be
``frozen in'', with strength proportional to the density of the field
lines, expanding with the plasma in the $yz$ plane. If the scale
factor in the $yz$ plane is given by $r$, it follows that the field
strength goes like $B \propto 1/r^2$.

Whilst homogeneity is preserved, the magnetic field breaks isotropy,
so the universe can be described using the Bianchi I model, with
metric 
\be\lab{metric4d} ds^2_4=\gamma_{\mu\nu}dx^\mu
dx^\nu=-d\tau^2+\lambda(\tau)^2 dx^2+r(\tau)^2(dy^2+dz^2).  
\ee 
Given that the only non-zero components of the field strength are
given by $F^y{}_z=-F^z{}_y=\sqrt{2} b/r(\tau)^2$, where $b$ is a
constant, one can easily check that Maxwell's equations are satisfied,
\be \nabla_\mu F^{\mu\nu}=0=\nabla_{[\alpha} F_{\mu\nu ]}. \ee The
remaining equations of motion are given by Einstein's equation with a
cosmological constant  \be\lab{eom}
G_{\mu\nu}+\Lambda_4\gamma_{\mu\nu}=8\pi G_4 T_{\mu\nu}, \ee where
$G_{\mu\nu}$ is the Einstein tensor and $T_{\mu\nu}$ is the energy
momentum tensor coming from the magnetic field. This is given by \be
\lab{Tmn} T^\mu_{\ \nu}=-\left(F^{\mu}{}_\gamma
F^\gamma{}_\nu-\frac{1}{4}\delta^\mu_{\ \nu} F^{\gamma}{}_\delta
F^\delta{}_\gamma\right)=\mathrm{diag}\left( -\frac{b^2}{r^4},
-\frac{b^2}{r^4}, \frac{b^2}{r^4}, \frac{b^2}{r^4} \right), \ee Note
that we could have obtained the same energy momentum tensor using an
electric field configuration with the same anisotropic
structure. However, as we have already mentioned, such a scenario
would not be physically relevant as the electric field would be
screened by the plasma in the early universe.

Provided $r(\tau)$ is not constant, it is convenient to treat
$\lambda$ as a function of $r$, so that Einstein's equations
(\ref{eom}) lead to the following 
\ba\lab{eom1}
\left(r^2(\lambda^2)''-2\lambda^2\right)r^2\Lambda_4 +\frac{8\pi G_4
b^2(r^2\lambda^2)''}{r^2}  &=&0, \\ \lab{eom2}
\dot{r}^2\left(r(\ln \lambda^2)'+1\right)-\Lambda_4 r^2
-\frac{ 8\pi G_4 b^2}{r^2} &=&0, \ea 
where
$'\equiv\partial_r$ and $\dot{}\equiv\partial_\tau $. Our way of
parametrising $\lambda$ is particularly useful for analysing the rate
of isotropization, since one could define cosmic time using $r$, as
opposed to $\tau$, and track the evolution of $\lambda$ using equation
(\ref{eom1}).  Indeed, although equation (\ref{eom1}) is non-linear,
it turns out that it has a simple analytic solution given by
\cite{exacsols,coles}  
\be \lab{lambda2} \lambda^2=
r^2-\frac{24\pi G_4 b^2}{\Lambda_4 r^2}+\frac{c}{r}, \ee
where $c$ is an 
integration constant. Plugging our solution back into (\ref{eom2}), one finds
that 
\be \lab{dotr}
\left(\frac{\dot r}{r}\right)^2=\frac{\Lambda_4}{3}
\frac{\lambda^2}{r^2}=\frac{\Lambda_4}{3}-\frac{8\pi G_4 b^2}{r^4},
\ee
where we have set the integration constant $c=0$ for
simplicity. Equation (\ref{dotr})
is consistent with our assumption that $r(\tau)$ is not
constant. Indeed, we can solve equation (\ref{dotr}) explicitly to give
\ba 
r(\tau) &=& \left(\frac{24\pi G_4
    b^2}{\Lambda_4}\right)^{\frac{1}{4}}\sqrt{\cosh\left(\sqrt{\frac{4\Lambda_4}{3}}(\t-\t_0)\right)}\lab{4dsola}, \\
\lambda(\tau) &=&  \left(\frac{24\pi G_4
    b^2}{\Lambda_4}\right)^\frac{1}{4}\frac{\sinh\left(\sqrt{\frac{4\Lambda_4}{3}}(\t-\t_0)\right)}{\sqrt{\cosh\left(\sqrt{\frac{4\Lambda_4}{3}}(\t-\t_0)\right)}}\lab{4dsolb},
\ea
where $\t_0$ is a constant. It is easy to see that the $yz$ plane expands quicker than the  $x$ direction. This is because the magnetic field exerts
negative pressure in the transverse
directions, leading to a greater expansion rate. If we consider photons propagating in such a background, those travelling in the $yz$ plane will be redshifted more than those travelling along the $x$ direction. In a more phenomenological scenario, this effect leads to temperature anisotropies in the CMB which can then be used to place bounds on the size of the magnetic field~\cite{CMBbound}.

We conclude this section by going back in time, and examining the behaviour close to the singularity, at which point $\lambda$ vanishes, while $r$ remains non-zero. Close to the singularity, our solution approaches a Milne universe, as we might
have expected from the ultralocal behaviour \cite{BKL}. The Milne
universe is a stable Kasner solution that does not undergo any Kasner
bounce or oscillation\footnote{It is a stable  solution
  only in the case of having an axial symmetry, otherwise any small
  fluctuations can lead to a complete different trajectory in the
  Kasner space, as one can see from the billiard picture of the
  Bianchi solutions (see for example \cite{damour}).}. If we consider $c \gtrsim 1$, the solution close to the singularity approaches the other Kasner solution given by $ds^2=-d\tau^2+\tau^{-2/3}dx^2+\tau^{4/3}(dy^2+dz^2)$. We can therefore think of $c$ as a measure of ``vacuum anisotropy'', breaking anisotropy even when there is no magnetic field. Later on  we will briefly discuss its role in the evolution of the
 anisotropy parameter.
\section{A magnetic brane in a boosted Schwarzschild bulk} \lab{sec:bw}
Whereas there is a wealth of literature discussing large scale
magnetic fields in standard four-dimensional General Relativity, there
is far less known about magnetic fields on Randall-Sundrum type
braneworlds. To our knowledge, existing studies (see, for
example,~\cite{magbrane, ecc, borniso, noiso}) have all been carried
without a complete description of the bulk geometry. In our view, this
is a dangerous game to play since there is no reason to believe that a
solution that is regular near the brane is also regular deep inside
the bulk. Indeed, it is well known that the effective gravitational
equations on the  brane are not closed, and that in general one must
first solve for the bulk before solving for the brane~\cite{sms}. This
is the approach we will take here.

In the single brane Randall-Sundrum scenario~\cite{rs}, our universe
corresponds to a 3-brane embedded in a five-dimensional anti-de Sitter
space. All standard model fields and interactions are confined to the
brane, although gravity is allowed to propagate through the bulk. The
bulk contains no energy-momentum other than a negative cosmological
constant, although matter excitations on the brane {\it can} alter the
geometry of the bulk. How would a braneworld magnetic field alter the
bulk geometry?

To get an idea of what happens, consider the case of  a cosmological
brane supporting a perfect fluid. A generalised form of Birkhoff's
theorem guarantees that the bulk geometry is a piece of $AdS_5$ or
Schwarzschild-$AdS_5$~\cite{bcg}. It has been argued that thermal
radiation of gravitons off the brane carries energy into the bulk,
unavoidably leading to the formation of the bulk black
hole~\cite{thermalgravitons}.  This will also be the case for a braneworld magnetic field, although isotropy will now be
broken on the brane. However, we might also consider the possibility that the graviton radiation will
not only carry energy into the bulk but also momentum, in equal and opposite amounts on either side of the brane. This
suggests that the bulk
may well be described by the planar limit of a Myers-Perry-$AdS_5$ solution~\cite{MP}.

 In five dimensions, a black hole may carry two independent angular
momenta $(J_1,J_2)$ in its two orthogonal planes of rotation, so that
the general solution is quite complicated. However, we will focus on a
special case for which the solution for a rotating black hole
simplifies considerably, as we shall now demonstrate. Consider first
the general Myers-Perry-$AdS_{5}$ solution~\cite{MP} with $J_1 \neq
J_2$, which has isometry group $\mathbb{R}\times U(1)^2$.  If one
takes equal angular momenta in the orthogonal planes, $J_1=J_2$, then
the isometry is enhanced\footnote{Note that for Schwarzschild, the
isometry group is $\mathbb{R}\times SU(2)\times SU(2)$.} to
$\mathbb{R}\times SU(2)\times U(1)$ . Let $r$ be the radial coordiante
of the black hole. The important property of the $J_1=J_2$ Myers-Perry
solution is that the surfaces of constant $r$ are {\it homogeneous}
spaces. While the event horizon of a static black hole is a round
$S^3$, the event horizon of the $J_1=J_2$ Myers-Perry is a
homogeneously squashed $S^3$. This makes the solution particularly
suitable for embedding homogenenous but anisotropic Bianchi
braneworlds.

In asymptotically anti-de Sitter space, the simplified Myers-Perry
solution can be written~\cite{MP}
\begin{equation} \lab{MPads}
ds^2 = -f^2dt^2 + g^2dr^2 + \frac{h^2}{4}(d\psi + \sin\theta d\phi -
\Omega dt)^2 + \frac{r^2}{4}(d\theta^2 + \cos^2\theta d\phi^2),
\end{equation}
where
\begin{eqnarray}\lab{func}
f ^2 & = & \frac{r^2}{g^2h^2}, \qquad h^2 = r^2\left(1 +
\frac{2ma^2}{r^4}\right), \qquad \Omega  =  \frac{4ma}{r^4 + 2ma^2},
\nonumber \\ g^{-2} & = & 1 + k^2r^2 - \frac{2m\Xi}{r^2} +
\frac{2ma^2}{r^4}, \qquad \Xi = 1 - k^2a^2,
\end{eqnarray}
 and the angles range over the following values:  $0 \leq \psi \leq 4\pi,  0 \leq \phi \leq  2\pi, 0 \leq
\theta \leq \pi$. Note that we have shifted $\theta \rightarrow \theta - \pi/2$ relative to
standard conventions for reasons that will become apparent shortly.  Provided $m \geq 0, ~\Xi>0$, there is a regular event horizon at the largest
root of $g^{-2}$. The Myers-Perry-$AdS_{5}$ solution satisfies the vacuum Einstein equations with negative cosmlogical constant and the Ricci tensor satisfies $R_{ab} =
-4k^2g_{ab}$ where $k$ is the inverse $AdS_{5}$ radius. We can relate the parameters
$(m,a)$ to the total energy and angular momenta of the
black hole via
\begin{equation}\label{bhenergy}
E = \frac{\pi m}{4G_5} \left(3 + a^2k^2\right), \qquad J_\psi = \frac{\pi
ma}{G_5}.
\end{equation} 
where $G_5$ is the five-dimensional Newton's constant. Since we will
be interested in Bianchi I braneworlds of the form (\ref{metric4d}),
it is convenient to consider the planar limit of the metric
(\ref{MPads}). This can be obtained by  `zooming' into the horizon as
follows. For constant $\epsilon$, we first redefine the coordinates
\begin{equation} \lab{limit}
t \to  \epsilon{{t}}, \qquad r \to \frac{{r}}{\epsilon}, \qquad  \theta
\to \epsilon{Y}, \qquad \psi\to  \epsilon{X}, \qquad \phi \to
\epsilon{{Z}},
\end{equation} 
and rescale $m \to  \epsilon^{-4} m$.  Now take the  limit $\epsilon
\to 0$, to give a two-parameter solution describing a planar
``black hole'', with a \emph{flat} $\mathbb{R}^3$
horizon\footnote{Strictly, the planar solution is not a black
hole since it has a non-compact horizon.}
\begin{equation} \lab{planarMP}
ds^2 = -f^2dt^2 + g^2dr^2 + \frac{h^2}{4}(dX - \Omega dt)^2 +
\frac{r^2}{4}(dY^2 +  dZ^2).
\end{equation}
The metric functions are the same as those given in~(\ref{func}), with
the exception of $g$, which now takes the form
\begin{equation}
 g^{-2} =  k^2r^2 - \frac{2m\Xi}{r^2}.
\end{equation} 
This solution is in fact planar Schwarzschild-$AdS_5$, written in a boosted coordinate system. To see this, consider the planar Schwarzschild-$AdS_5$ metric
\begin{equation} \lab{schw}
ds^2 = -V(r)d\hat t^2 + \frac{dr^2}{V(r)} +
r^2(d\hat X^2+d\hat Y^2 +  d\hat Z^2), \qquad V(r)=k^2r^2-\frac{2m\Xi}{r^2}.
\end{equation}
and transform to a boosted coordinate system, given by
\be
\hat t =\frac{1}{\Xi} \left(t+a\frac{X}{2} \right), \qquad \hat X=\frac{1}{\Xi} \left(\frac{X}{2}+k^2a t\right), \qquad \hat Y=\frac{Y}{2}, \qquad \hat Z=\frac{Z}{2} .
\ee
An observer at $r \to \infty$ using Schwarzschild time, $\hat t$, will measure the following bulk energy and momentum, 
\be
\hat E=\frac{3 m\Xi}{8 \pi G_5}\int d\hat Xd\hat Y d \hat Z, \qquad \hat P_{\hat X}=\hat P_{\hat Y}=\hat P_{\hat Z}=0.
\ee   
However, the boosted observer, using time, $t$, will measure different energy and momentum, which can be easily calculated using either background subtraction~\cite{BY} or the boundary counterterm method~\cite{BK}, to give
\be
E= \frac{m}{64 \pi G_5} (3+a^2k^2) \int dX dY dZ, \qquad P_X=\frac{ma}{16\pi G_5}\int dX dY dZ, \qquad P_Y=P_Z=0 .
\ee
Notice the similarity with the original expressions (\ref{bhenergy}) for the energy and angular momentum of the Myers-Perry solution. Although these expressions are clearly divergent due to the infinite volume integral, we will eventually make use of the ``densities'', ${\cal E}=E/ \int dX dY dZ$ and ${\cal P}_X =P_X/\int dX dY dZ $, which, of course, remain finite and well defined. The key point to note is that the boosted observer actually measures non-zero momentum along the $X$-direction. We expect the same to be true for an observer on the brane, since the presence of the magnetic field, and the breaking of isotropy leads to equal and opposite momentum being carried into the bulk by hot gravitons.
\subsection{Embedding the brane}
We now turn our attention to constructing the brane. We will do this
in the usual way, cutting and pasting together two $5D$ spacetimes,
$\mathcal{M}_+$ and $\mathcal{M}_-$, in such a way that the Israel
junction conditions~\cite{junc} are satisfied across the brane,
$\Sigma=\del \mathcal{M}_\pm$ (see, for example,~\cite{bcg}). Although
it is straightforward to embed a homogeneous and isotropic brane in a
static Schwarzschild-AdS bulk~\cite{bcg}, the situation for an anisotropic brane is obviously more complicated. It turns out that these difficulties can be alleviated somewhat by working in a boosted coordinate system, as opposed to global coordinates in the bulk. For this reason we have chosen $\mathcal{M}_+$
and $\mathcal{M}_-$ to correspond to the simplified planar solution
(\ref{planarMP}) but with {\it equal and opposite
momentum}, $P_X^+=-P_X^-$. In $\mathcal{M}_\pm$, the metric is therefore given by
\begin{equation} \lab{planarMP1}
ds^2 =g^{\pm}_{ab}dX^a dX^b= -f^2dt^2 + g^2dr^2 + \frac{h^2}{4}(dX \mp
\Omega dt)^2 + \frac{r^2}{4}(dY^2 +  dZ^2).
\end{equation} 
We require the bulk momentum to flip sign because we will be
constructing a non-rotating Bianchi I brane. This makes sense when one
considers gravitons carrying thermal radiation into the bulk. By
conservation of momentum, a graviton carrying positive 
momentum into $\mathcal{M}_+$ must be compensated for by a graviton
carrying negative  momentum into $\mathcal{M}_-$. 

On each side of the bulk, the brane
can be regarded as a boundary surface, $\Sigma=\del \mathcal{M}_\pm$, described
by the following embedding,
\begin{equation}
t = t(\tau), \qquad r = r(\tau), \qquad X = x  \pm \int^{\tau}
\Omega(r(\tau'))\dot{t}(\tau') d\tau', \qquad Y=y, \qquad Z=z,
\end{equation}
where  $t(\tau)$ and $r(\tau)$ are chosen so that
\be \lab{constraint}
-f^2\dot{t}^2 + g^2\dot{r}^2 = -1.
\ee
In both cases the bulk geometry corresponds to  $0<r<r(\tau)$, which means we retain the
black hole, as opposed to the asymptotic region. The induced metric on
the brane now takes the Bianchi I form
\begin{equation} \lab{brmetric}
ds^2_{b} =\gamma_\mn dx^\mu dx^\nu=-d \tau^2+
\frac{h(r(\tau))^2}{4}dx^2 +
\frac{r(\tau)^2}{4}(dy^2+dz^2).
\end{equation}
Let us suppose that the brane has tension, $\sigma$, and supports a
magnetic field with energy momentum tensor (\ref{Tmn}).  The Israel
equations~\cite{junc} require that 
\be \lab{israel}
K^+_\mn-K^+ \ga_\mn+K^-_\mn-K^-
\ga_\mn=8 \pi G_5 (-\sigma\gamma_\mn+T_\mn),
\ee 
where $K^\pm_\mn$ is the
extrinsic curvature of the brane in $\mathcal{M}_\pm$. We define  it
as $K_\mn=\frac{1}{2}\mathcal{L}_n \ga_\mn$, the Lie derivative of the
induced metric on $\Sigma =\del \mathcal{M}_\pm$ with respect to  the
normal pointing {\it out of} $\mathcal{M}_\pm$.  Note that our
convention differs slightly from the one used in~\cite{junc} in that
we have taken the normal to flip direction as we cross the brane. 

The off-diagonal components of the Israel equations (\ref{israel}) vanish identically on account of the two
bulk black hole having equal and opposite  momenta. Although
the diagonal components are non-zero, given the constraint
(\ref{constraint}),  it turns out there are only two more independent
equations of motion:
\begin{eqnarray}
\frac{2 f \dot t}{g}\del_r\ln(hr^2)&=&-8\pi G_5
\left(\sigma+\frac{b^2}{r^4}\right), \lab{is1} \\
\frac{2 f \dot t}{g}\del_r\ln(h/r)&=&16\pi G_5
\frac{b^2}{r^4} \lab{is2}.
\end{eqnarray}
These two equations are consistent if and only if $a^2=3b^2/2m\sigma$,
illustrating explicitly how a magnetic field on the brane must be
supported by momentum in the bulk. Combining equations
(\ref{constraint}), (\ref{is1}) and (\ref{is2}), we find that
\be
\left(\frac{\dot r}{r}\right)^2=\left(\frac{4 \pi G_5 \sigma h^2}{3r^2}\right)^2-\frac{1}{r^2g^2}.
\ee
When the magnetic field is absent ($b=0$), note that the the system
reduces to a homogeneous and isotropic brane embedded in unboosted $AdS_5$ or
(planar) 
Schwarzschild-$AdS_5$ bulk
as one might have expected~\cite{rs,bwholo3}.  One can easily check
that the brane cosmological
constant can be given in terms of the tension and the bulk
cosmological constant
\be \lab{Lam4}
\Lambda_4=3\left[\left(\frac{4 \pi G_5 \sigma }{3}\right)^2-k^2\right],
\ee
in agreement with the standard result (see, for example,
~\cite{rs, bcg,bwholo3}). If $m \neq 0$, the black hole mass is seen
as dark radiation by an observer on the brane~\cite{bwholo3}. When we switch on the magnetic field ($b
\neq 0)$,
gravitons carry momentum into the bulk ($ma \neq 0$), and the brane turns
into a Bianchi I universe. We can compare this directly with the Bianchi I
universe derived in the previous section for standard $4D$ GR by
transforming coordinates $x \to 2 x$, $y \to 2y$ and $z \to 2z$. Then the brane
metric (\ref{brmetric}) takes the form (\ref{metric4d}) but with
\be \lab{bw1}
\lambda^2=h^2=r^2+\frac{3b^2}{\sigma r^2},
\ee 
and
\be \lab{bw2}
\left(\frac{\dot r}{r}\right)^2=\left(\frac{4 \pi G_5 \sigma \lambda^2}{3r^2}\right)^2-k^2
\frac{\lambda^2}{r^2}+\frac{2m}{r^4}.
\ee
Now Newton's constant on the brane is given in terms of the bulk
Newton's constant as follows~\cite{rs, bcg,bwholo3}
\be \lab{Gn}
G_4=G_5 \left(\frac{4 \pi G_5 \sigma}{3}\right).
\ee
Using (\ref{Lam4}) and (\ref{Gn}), we can rewrite equations (\ref{bw1})
and (\ref{bw2}) in terms of $4D$ quantities,
\ba
\lambda^2&=&r^2+\frac{12\pi G_4 b^2}{(\Lambda_4+3k^2)r^2}, \lab{lam2}\\
\left(\frac{\dot r}{r}\right)^2 &=& \frac{\Lambda_4}{3}+
\left(\frac{2\Lambda_4+3k^2}{\Lambda_4+3k^2}\right)\frac{4\pi G_4
  b^2}{r^4}+\frac{2m}{r^4}+\frac{3}{\Lambda_4+3k^2}\left(\frac{4\pi
  G_4 b^2}{r^4}\right)^2 \lab{Hr}.
\ea
Of course, $k$ measures the bulk AdS curvature, and is
not a $4D$ quantity. However, we can use the AdS/CFT correspondence to
relate it to the effective number of degrees of freedom in the dual
CFT~\cite{bwholo3, adscft, bwholo1, bwholo2}
\be
g_* \sim N^2 \sim \left(\frac{1}{kl_4}\right)^2,
\ee
where $l_4 \sim \sqrt{\hbar G_4}$ is the $4D$ Planck length. We shall emphasize the role played by the dual conformal field theory by calculating the Einstein tensor on the brane
\be
G_\mn(\ga)=-\Lambda_4 \ga_\mn+8 \pi G_4 \left[T_\mn+
\Delta T_\mn \right].
\ee 
The braneworld corrections are all included in $\Delta T_\mn$, which is given by 
\be
\Delta T^\mu_\nu =\textrm{diag} (-\Delta\rho, \Delta p_{x},\Delta p_y, \Delta p_z ),
\ee
where
\ba
\Delta \rho&=&\frac{m(3+k^2a^2)}{4\pi G_4r^4}+\mathcal{O}(\rho_B^2) \lab{Deltarho},\\ 
\Delta p_x&=&\frac{m\left(1+\left(3k^2+\frac{4\Lambda_4}{3}\right)a^2\right)}{4\pi G_4r^4}+\mathcal{O}(\rho_B^2) \lab{px},\\ 
\Delta p_y&=& \Delta p_z=\frac{m\left(1-\left(k^2+\frac{2\Lambda_4}{3}\right)a^2\right)}{4\pi G_4r^4}+\mathcal{O}(\rho_B^2) \lab{py},
\ea
 with $\rho_B=b^2/r^4$. Now for large $r$, we interpret the CFT  energy as being the energy of the bulk as measured by an observer on the brane. This goes like~\cite{bwholo3}
\be
E_\tt{CFT}\approx 2E \left(\frac{k}{4\pi G_5 \sigma/3}\right)^2 \dot t \sim \frac{m(3+k^2a^2)}{32\pi G_4r } \int dXdYdZ,
\ee
where we have used equation (\ref{Gn}) and the fact that $\dot t \sim 4\pi G_5 \sigma/3k^2r$. Given that the volume of the brane, $V \sim r^3\int dx dy dz =(r/2)^3  \int dXdYdZ$, we conclude that the CFT energy density is given by
\be \lab{rhoCFT}
\rho_\tt{CFT}=\frac{E_\tt{CFT}}{V} \approx \Delta \rho.
\ee
We now see explicitly why the braneworld corrections, $\Delta T_\mn$, can be associated with the
CFT stress energy tensor, $\langle
T^\textrm{CFT}_\mn \rangle$, at least to leading order~\cite{bwholo3,
  adscft, bwholo1, bwholo2}.

Because  gravity is localised in the Randall-Sundrum model at
distances $d \gg k^{-1}$~\cite{rs}, we might naively expect the CFT correction
to be suppressed on large enough scales.  In other words, the energy
density of the CFT, $\rho_{CFT} \ll \rho_B$ whenever $8 \pi
G_4 \rho_B \ll k$. However, recall 
the condition $\Xi=1-k^2a^2>0$, required in order to avoid a
naked singularity in the bulk. Given the fact that
$a^2=3b^2/2m\sigma$, and assuming $\Lambda_4 \ll k^2$, this condition
translates into the an approximate bound on the energy density of the
magnetic field,
\be \lab{rhoBbound}
\rho_B \lesssim
\rho_{CFT}.
\ee
It is clear that we can never suppress the energy density
of the CFT relative to the magnetic field in this construction. We do
not know whether this is merely an artifact of requiring a Schwarzschild-AdS bulk, or whether it represents something more generic, at least when
non-perturbative effects are fully taken into account.

The CFT stress energy arises from particle production, exerting
additional negative pressure along the $x$
direction. This effect is so marked that the $x$ axis actually expands
faster than the $yz$ plane.  This is in complete contrast with the
standard $4D$ picture in which the reverse is true, and reflects the
fact CFT corrections can never be suppressed on any scale. If this
effect is generic, and 
also applied to more phenomenological branes worlds (eg a Bianchi
VIIh brane), it would give a potential test of the Randall-Sundrum
scenario.  At present gamma-ray bursts provide the most promising
means of detecting a primordial magnetic 
field~\cite{detect}. If we suppose that a homogeneous
field is detected, and 
seen to point along a certain direction, then in standard $4D$ GR, one
would expect photons travelling from  that direction to be redshifted less than
those of travelling along the orthogonal direction. As we have just seen, the opposite would be
true in a braneworld scenario, so this could indeed be a potentially useful test of
large extra dimensions. The production of CFT particles also has the effect of damping the
overall anisotropy. We will analyse this in more detail in the next
section.
\section{Evidence for anisotropy dissipation in a braneworld}
\lab{sec:compare}
In the previous section, we considered the case of this Universe corresponding to a Bianchi I brane (\ref{metric4d}),  embedded in between two Schwarzschild-AdS bulk spacetimes, boosted equal amounts in opposite directions. We would like to study the anisotropic evolution of this solution as we adjust the various parameters, at the same time comparing them  with the standard $4D$ scenario discussed in section~\ref{sec:4dgr}. To this end, it is convenient to define the
following tensor on surfaces of constant $\tau$:
\be
\Theta^i{}_j=\textrm{diag}\left(\frac{\dot
\lambda}{\lambda}, \frac{\dot r}{r}, \frac{\dot r}{r} \right). \ee 
This
measures the velocity of such surfaces, corresponding to the extrinsic
curvature in the absence of vorticity (for more detail, see, for
example,~\cite{BMT}). The trace now gives the volume expansion rate,
\be \Theta=\frac{\dot \lambda}{\lambda}
+2\frac{\dot r}{r}  \ee and whereas the traceless piece gives
the the shear tensor \be \sigma^i{}_j
=\Theta^i{}_j-\frac{1}{3}\Theta\delta^i_j=\left(\frac{\dot
\lambda}{\lambda}-\frac{\dot
r}{r}\right)\textrm{diag}\left(\frac{2}{3}, -\frac{1}{3}, -\frac{1}{3}
\right). \ee  
A useful measure of the overall anisotropy is given by
the ratio, $\Sigma\equiv\sigma/H$,  where the shear scalar,
$\sigma=\sqrt{\sigma^i{}_j\sigma^j{}_i/2}$ and the mean expansion
rate, $H=\Theta/3$.  It is straightforward to see that
$\Sigma=3/\sqrt{2}$ for the Milne Universe and $\Sigma=0$ for any
isotropic model. 
\begin{figure}[t!]
\begin{center}
\resizebox*{4.8in}{3in}
{\includegraphics{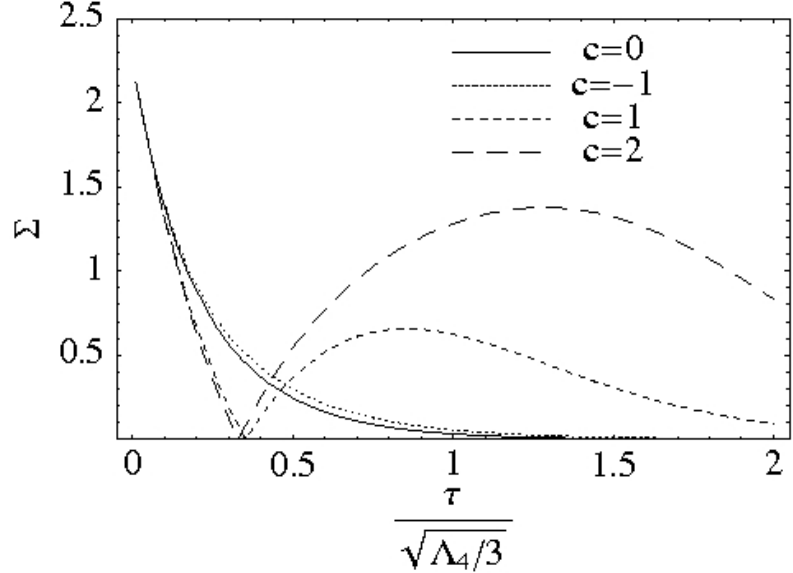}}
\caption{Plots show the evolution over time of the anisotropy parameter, $\Sigma$, in the $4D$ case, for various values of the vacuum anisotropy component, $c$, and all other parameters held fixed.}
\label{fig:4dcompare}
\end{center}
\end{figure}
Let us consider for the moment the standard $4D$ solutions discussed in section~\ref{sec:4dgr}. In the absence of any vacuum anisotropy ($c=0$), we saw how  the solution (\ref{4dsola}) and(\ref{4dsolb}) approached a Milne Universe close to the singularity, which means it  started out with $\Sigma=3/\sqrt{2}$. The solution then decreases continuously to
zero, as expected since the cosmological constant will ultimately
take over the evolution and restore isotropy at late times~\cite{wald}. This behaviour is shown explicitly in Fig.~\ref{fig:4dcompare}. In  contrast, if enough vacuum anisotropy is present ($c \gtrsim 1$), $\Sigma$ will always undergo an
increasing phase, as also shown in Fig.~\ref{fig:4dcompare},  corresponding to an early era of ``vacuum anisotropy domination''. We will not pursue this particular solution any further, but it may be an interesting topic for future research in its own right.

In the remainder of this section, we will focus on comparing the evolution of anisotropy in various brane scenarios and in the standard $4D$ case with no  vacuum anisotropy. There are some obvious differences. In the standard $4D$ case described by equations (\ref{4dsola}) and (\ref{4dsolb}), the anisotropic
parameter depends only on $\Lambda_4$ and the energy density of the magnetic field $\rho_B=b^2/r_*^4$ at a
given time, $\tau=\tau_*$. In contrast, the braneworld solution, also depends on the five
dimensional parameters $m$ and $k$. These have a four-dimensional
interpretation in terms of the temperature and the effective number of
degrees of freedom of the CFT respectively~\cite{bwholo3, adscft,
  bwholo1, bwholo2}, and contribute to the CFT energy density given by
equations (\ref{Deltarho}) and  (\ref{rhoCFT}). 

\begin{figure}[t!] 
\begin{center}
\resizebox*{4.8in}{3in}
{\includegraphics{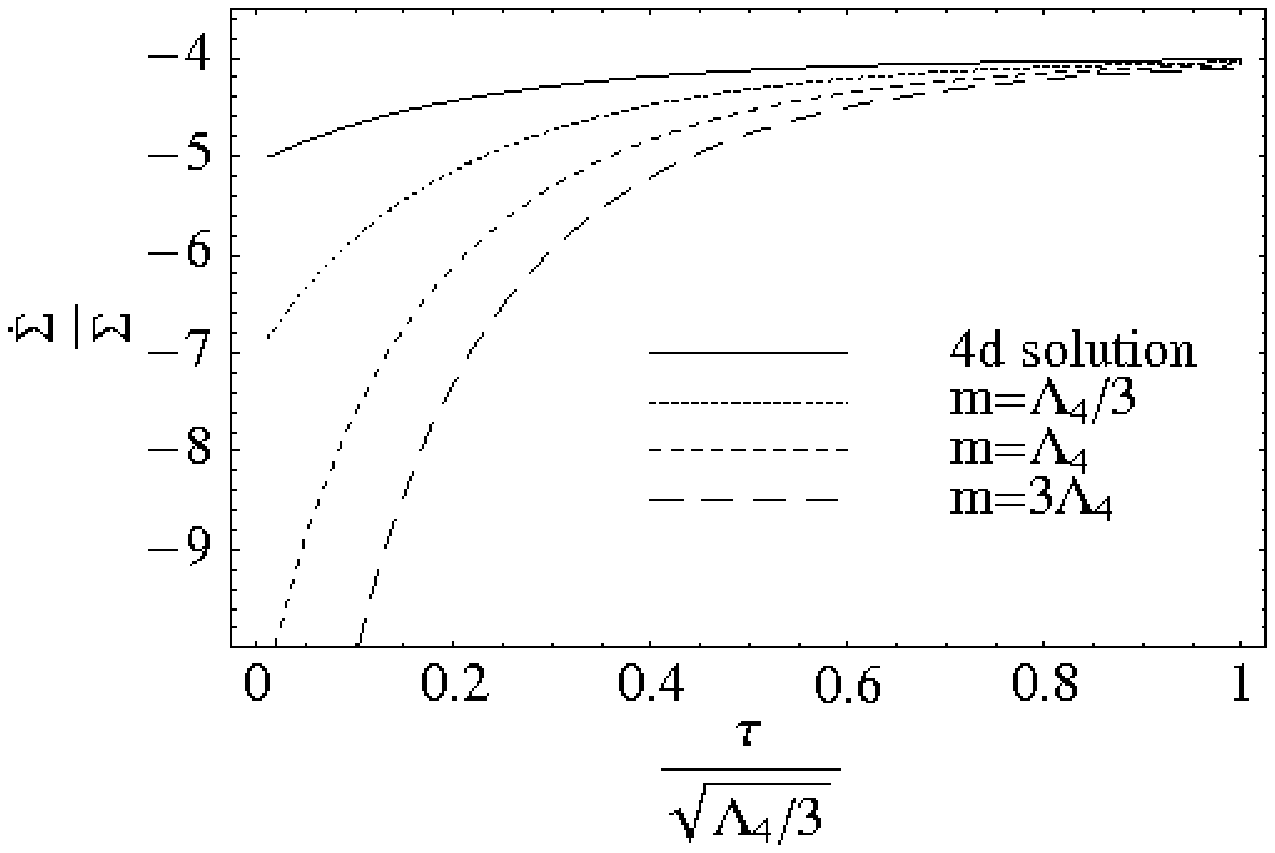}}
\resizebox*{4.8in}{3in}
{\includegraphics{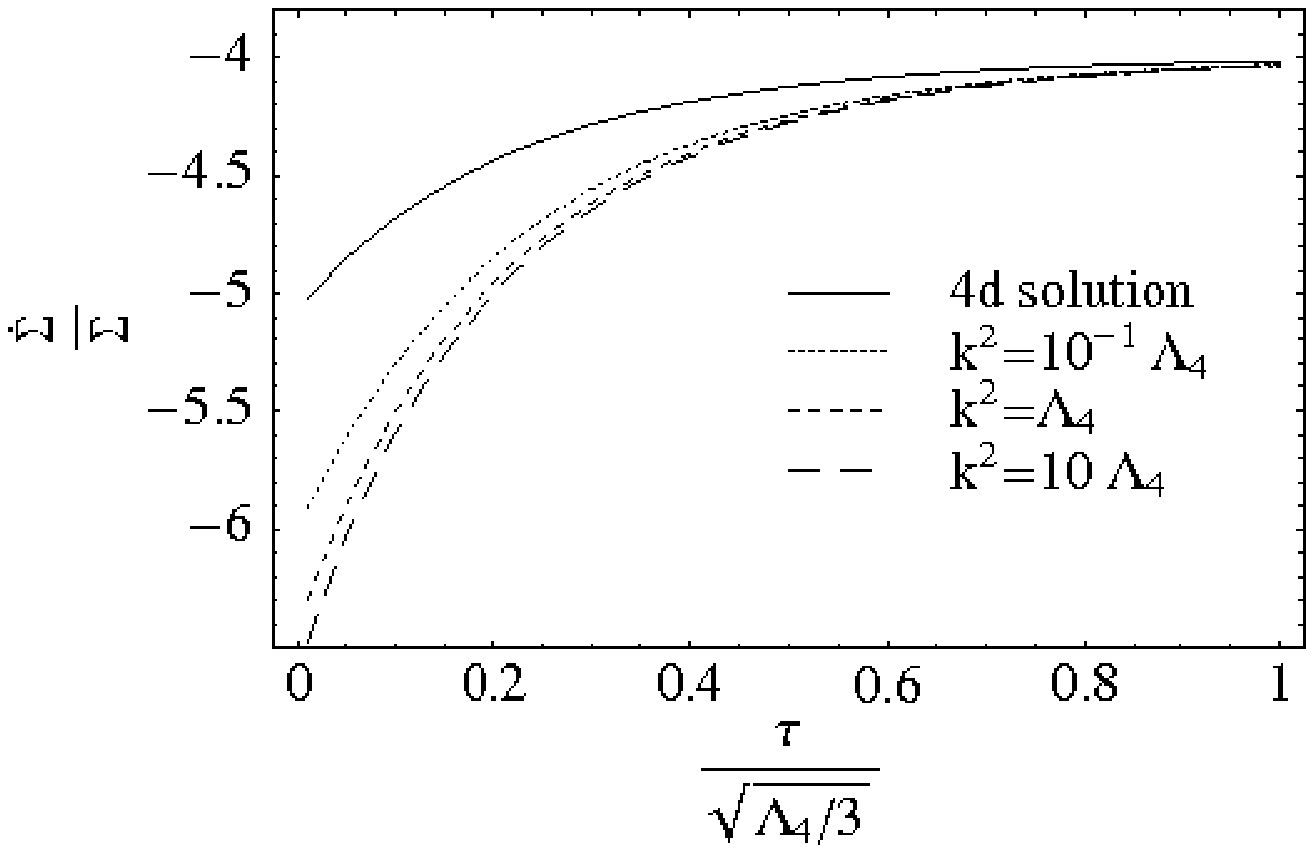}} 
\caption{Plots show the rate of change of the anistropy parameter
  $\dot{\Sigma}/\Sigma$ as a function of time (in units of a de Sitter Hubble
  radius for a cosmological constant $\Lambda_4$). All the $4D$
  parameters are held fixed (\ie $\Lambda_4$, $G_4$ and
  $b^2=10^{-1}\Lambda_4/(8\pi G_4)$). The different curves show how
  the CFT affects the early universe isotropization as a function of
  $m$ for fixed $k$ (with $k^2=\Lambda_4$) in the top plot and as a
  function of $k$ for fixed $m$ (with 
  $m=\Lambda_4/4$) in the bottom plot. In all cases
  we have used a normalization of the scale factor given by $r(\tau_*)=1$
  with $\tau_*=\frac{1}{3}\times 10^{-2}\Lambda_4$.}
 \label{plots}
\end{center}
\end{figure}

In all cases, the cosmological constant term will eventually dominate, and all the anisotropy will eventually be dissipated away~\cite{wald}. However, we are interested in the {\it rate} of dissipation, {\it before} the $\Lambda_4$ term starts to play a significant role. For this reason, we have chosen to plot  the averaged rate of change, $\dot{\Sigma}/{\Sigma}$,
 as a function of proper time, over a period $\Delta \tau \sim \sqrt{\Lambda_4/3}$ (see Fig. \ref{plots}). 
We see that in all cases  $\dot \Sigma/ \Sigma$ is more negative for the braneworld than in  standard $4D$ GR.  This means that the isotropization mechanism is indeed {\it more} efficient on a Randall-Sundrum brane, a phenomenon we will dub {\it braneworld isotropization}, and which we attribute to particle production in the dual CFT. A closer look at Fig.~\ref{plots} reveals that braneworld isotropization becomes more efficient for larger values of $m$ or $k$ (with all the other parameters
held fixed). To see why this makes sense from the CFT perspective, consider again the first order contribution to
the CFT energy momentum tensor given by equations (\ref{Deltarho}) to (\ref{py}),
\be
\langle T^\textrm{CFT}_\mn \rangle =\Delta T_\mn=T^{(m)}_\mn+T^{(k)}_\mn,
\ee
where, using the relation $a^2=2b^2/2m\sigma$, and equation (\ref{Lam4}), we have separated out the $m$ and $k$ dependent contribution like so 
\ba
&&T^{(m)}{}^\mu_\nu= \frac{m}{4\pi G_4 r^4}\mathrm{diag}(-3, 1, 1, 1), \\
&&T^{(k)}{}^\mu_\nu= \rho_B\mathrm{diag}\left(-\frac{k^2}{2k^2+2\Lambda_4/3}, \frac{3k^2+4\Lambda_4/3}{2k^2+2\Lambda_4/3},  -\frac{k^2+2\Lambda_4/3}{2k^2+2\Lambda_4/3},  -\frac{k^2+2\Lambda_4/3}{2k^2+2\Lambda_4/3}\right) .
\ea
The first thing to note is that the $m$ dependent piece, $T^{(m)}_\mn$, is isotropic. This corresponds to the isotropic thermal radiation produced by the finite temperature CFT~\cite{bwholo3}. Clearly, when $m$ is large, the CFT is very hot and a lot of this isotropic radiation is produced. It soon dominates over the anisotropic components of energy-momentum, and anisotropy is quickly dissipated. In contrast, the $k$ dependent piece, $T^{(k)}_\mn$, is anisotropic. This represents the response of the CFT to the anisotropy created by the matter energy-momentum tensor (\ref{Tmn}).  Because the dual CFT is coupled to gravity in the Randall-Sundrum picture, the anisotropy leads to CFT particle production. We see explicitly how these particles exert positive pressure along the $x$-direction, in order to compensate for the corresponding negative pressure exerted by the magnetic field. To get an idea of the net effect, we combine the two competing anisotropic contributions
\ba
T^\tt{(aniso)}{}^\mu_\nu&=&T^\mu_\nu+T^{(k)}{}^\mu_\nu, \nonumber
\\
&=&\rho_B\mathrm{diag}\left(-\frac{3k^2+2\Lambda_4/3}{2k^2+2\Lambda_4/3}, \frac{k^2+2\Lambda_4/3}{2k^2+2\Lambda_4/3},  \frac{k^2}{2k^2+2\Lambda_4/3},  \frac{k^2}{2k^2+2\Lambda_4/3}\right).
\ea
For large $k$, it is easy to check that the net effect is actually isotropic! Naively, we might have expected the large $k$ limit to coincide with the $4D$ case. However, as we discussed at the end of the last section, one cannot consistently consider $8\pi G_4 \rho_B \ll k$, and at the same time switch off the contribution of the CFT. The heuristic arguments presented here are a little crude since they ignore the evolution of the scale factor $r(\tau)$, which also depends on $m$ and $k$, according to equation (\ref{Hr}). Nevertheless, they certainly give us a rough understanding of what is going on, and in any case, the plots given in Fig. \ref{plots} clearly demonstrate braneworld isotropization in action.
\section{Discussion} \lab{sec:discuss}
In this paper, we have explicitly demonstrated the existence of a new phenomenon, called {\it braneworld isotropization}, by means of a concrete example. We were able to embed a magnetic Bianchi I braneworld in between two Schwarzschild-$AdS_5$ spacetimes, boosted equal amounts in opposite directions. The magnetic field breaks isotropy on the brane, and leads to the production of thermal graviton radiation, that can carry energy-momentum into the bulk. The boosted coordinates mean that the brane observer sees equal and opposite momentum carried into the bulk by the  hot gravitons leaving the brane. We can view braneworld isotropization as anisotropic energy on the brane being dumped into the bulk.  From a completely $4D$ point of view, we can understand this effect using the AdS/CFT correspondence. In the Randall-Sundrum scenario, the gravitational physics in the bulk is equivalent to a strongly coupled CFT, cut-off in the UV, and minimally coupled to gravity on the brane. When isotropy is broken, the coupling of the CFT to gravity leads to particle production, the required  energy being drawn from the anisotropy.

This phenomenon can now be used to readily explain a number of existing results in the literature~\cite{maar1, maar2, orsaylot, frolov}. For example, an  anisotropic brane cannot support a perfect fluid in a static bulk~\cite{orsaylot},  because the leaking of anisotropy off the brane prevents the bulk from being static, at least when viewed by an asymptotic braneworld observer.  Note that in the solution described in this paper, we do not have a perfect fluid on the brane, and although the bulk is static according to an asymptotic observer using Schwarzschild time, it is {\it not} static according to an asymptotic observer using the boosted time coordinate. It is the latter time coordinate that is used by the asymptotic braneworld observer at large  $r(\tau)$. 

It is worth noting that we have been able to prove that the solution presented here, along with the known isotropic solutions~\cite{bcg}, are the most general solutions for a Bianchi I brane of the form (\ref{metric4d}), embedded in Schwarzschild-$AdS_5$ with some form of ``sensible'' matter on the brane. We have not included the proof since it is lengthy, and not particularly illuminating. However, it does tell us that we will have to work a bit harder if we want to, say, include an arbitrary perfect fluid on the brane in addition to the magnetic field. This is a subject for future study. Nevertheless, we anticipate that many of the features seen here will remain, in particular, momentum being generated in the bulk by the magnetic field on the brane, and particle production in the CFT leading to anisotropy dissipation.

Perhaps one of the most interesting features of this work was the possibility that one could detect a braneworld signature in the sky, as discussed near the end of section~\ref{sec:bw}. In the braneworld picture, particles are actually {\it over}-produced along the direction of the magnetic field, so that the expansion along that direction is {\it slower} than in the orthogonal directions, in complete contrast with what happens in standard $4D$ General Relativity. For this reason it would be interesting to extend this work to other, more phenomenological, Bianchi models. If, say, one of the spatial directions were  periodic, we might expect {\it angular} momentum to be carried into the bulk, and so we would look for an embedding in the general Myers-Perry spacetimes~\cite{MP, HHT}.  A perturbative study along these lines has been considered~\cite{Guth}, although a far more complete analysis is clearly required.

In the long term, we would like to consider  a physically realistic
scenario with a Bianchi VIIh brane containing a homogeneous magnetic
field along with an isotropic perfect fluid made up of matter,
radiation and a cosmological constant. Of course, this would require
an amalgamation of the ideas outlined in the previous two paragraphs.
It would be very interesting to see what effect braneworld
isotropization has on the CMB, and in particular the bounds on the
size of the primordial magnetic field~\cite{CMBbound}.  We must also
keep in mind constraints coming from nucleosynthesis. For example, if
braneworld isotropization is {\it too} efficient, then we might worry
that CFT particle production will result in too much dark radiation, leading to an unacceptably low baryon density parameter. 
\section*{Acknowledgements}
We would like to thank Enric Verdaguer, Ed Copeland, Paul Saffin,
Peter Coles, Nemanja Kaloper, Enrico Ramirez Ruiz, Roy
Maartens, Christos Charmousis, Fernando Torres Sanz, John Barrow, and Sigbjorn Hervik for useful discussion and correspondence. HKK is supported by the SFTC and thanks the Particle Theory Group, University of Nottingham, for their hospitality.

\end{document}